\documentclass[iop]{emulateapj}

\def\kms{km\,s$^{-1}$}
\def\dlambda{$\lambda\lambda$}

\slugcomment{Accepted to The Astrophysical Journal Letters on January
  4, 2014}

\shorttitle{SN Interaction with a Carrier of DIBs}
\shortauthors{Milisavljevic et al.}

\begin{document}

\def\cfa{1}
\def\yale{2}
\def\dartmouth{3}
\def\lco{4}
\def\salt{5}
\def\stsci{6}
\def\gsfc{7}
\def\berk{8}
\def\ut{9}
\def\ari{10}
\def\mp{11}
\def\inafpad{12}
\def\kyoto{13}
\def\uu{14}
\def\usz{15}

\title{Interaction Between The Broad-lined Type I\lowercase{c} Supernova 2012\lowercase{ap}\\
and Carriers of Diffuse Interstellar Bands}

\author{Dan Milisavljevic\altaffilmark{\cfa,$\dagger$},
        Raffaella~Margutti\altaffilmark{\cfa},
        Kyle~N.~Crabtree\altaffilmark{\cfa},
        Jonathan~B.~Foster\altaffilmark{\yale},
        Alicia~M.~Soderberg\altaffilmark{\cfa},                
        Robert~A.~Fesen\altaffilmark{\dartmouth},
        Jerod~T.~Parrent\altaffilmark{\dartmouth,\lco},
        Nathan~E.~Sanders\altaffilmark{\cfa},
        Maria~R.~Drout\altaffilmark{\cfa},
        Atish~Kamble\altaffilmark{\cfa},
        Sayan~Chakraborti\altaffilmark{\cfa},    
        Timothy~E.~Pickering\altaffilmark{\salt,\stsci},
        S.~Bradley~Cenko\altaffilmark{\gsfc,\berk},
        Jeffrey~M.~Silverman\altaffilmark{\ut},
        Alexei~V.~Filippenko\altaffilmark{\berk},       
        Robert~P.~Kirshner\altaffilmark{\cfa},
        Paolo~Mazzali\altaffilmark{\ari,\mp,\inafpad},
        Keiichi~Maeda\altaffilmark{\kyoto,\uu},\\
        G.~H.~Marion\altaffilmark{\ut},
        Jozsef~Vinko\altaffilmark{\ut,\usz}, and
        J.~Craig~Wheeler\altaffilmark{\ut}       
}

\altaffiltext{\cfa}{Harvard-Smithsonian Center for Astrophysics, 60
  Garden St., Cambridge, MA 02138}
\altaffiltext{\yale}{Yale Center for Astronomy and Astrophysics, Yale
University, New Haven, CT 06520, USA}
\altaffiltext{\dartmouth}{Department of Physics \& Astronomy, Dartmouth
                 College, 6127 Wilder Lab, Hanover, NH 03755, USA}
\altaffiltext{\lco}{Las Cumbres Observatory Global Telescope Network, Goleta, CA, USA}
\altaffiltext{\salt}{Southern African Large Telescope, PO Box 9,
Observatory 7935, Cape Town, South Africa}
\altaffiltext{\stsci}{Space Telescope Science Institute, 3700 San
  Martin Drive, Baltimore, Maryland 21218, USA}
\altaffiltext{\gsfc}{Astrophysics Science Division, NASA Goddard Space
  Flight Center, Mail Code 661, Greenbelt, MD 20771, USA}
\altaffiltext{\berk}{ Department of Astronomy, University of
  California, Berkeley, CA 94720-3411, USA}
\altaffiltext{\ut}{University of Texas at Austin, 1 University Station
  C1400, Austin, TX, 78712-0259, USA}
\altaffiltext{\ari}{Astrophysics Research Institute, Liverpool John Moores University,
Liverpool L3 5RF, United Kingdom}
\altaffiltext{\mp}{Max-Planck-Institut f\"ur Astrophysik,
  Karl-Schwarzschild-Strasse 1, 85748 Garching, Germany}
\altaffiltext{\inafpad}{INAF - Osservatorio Astronomico di Padova,
  Vicolo dell'Osservatorio 5, I-35122, Padova, Italy}
\altaffiltext{\kyoto}{Department of Astronomy, Kyoto University
Kitashirakawa-Oiwake-cho, Sakyo-ku, Kyoto 606-8502, Japan}
\altaffiltext{\uu}{Kavli Institute for the Physics and Mathematics
of the Universe (WPI), Todai Institutes for Advanced Study,
University of Tokyo, 5-1-5 Kashiwanoha, Kashiwa, Chiba 277-8583, Japan}
\altaffiltext{\usz}{Department of Optics and Quantum Electronics,
University of Szeged, Domter 9, 6720, Szeged, Hungary}
\altaffiltext{$\dagger$}{email: dmilisav@cfa.harvard.edu}

\begin{abstract}

  The diffuse interstellar bands (DIBs) are absorption features
  observed in optical and near-infrared spectra that are thought to be
  associated with carbon-rich polyatomic molecules in interstellar
  gas. However, because the central wavelengths of these bands do not
  correspond with electronic transitions of any known atomic or
  molecular species, their nature has remained uncertain since their
  discovery almost a century ago. Here we report on unusually strong
  DIBs in optical spectra of the broad-lined Type Ic supernova
  SN\,2012ap that exhibit changes in equivalent width over short ($\la
  30$ days) timescales. The 4428\,\AA\ and 6283\,\AA\ DIB features get
  weaker with time, whereas the 5780\,\AA\ feature shows a marginal
  increase. These nonuniform changes suggest that the supernova is
  interacting with a nearby source of the DIBs and that the DIB
  carriers possess high ionization potentials, such as small cations
  or charged fullerenes.  We conclude that moderate-resolution spectra
  of supernovae with DIB absorptions obtained within weeks of outburst
  could reveal unique information about the mass-loss environment of
  their progenitor systems and provide new constraints on the
  properties of DIB carriers.

\end{abstract}

\keywords{astrochemistry --- molecular processes --- ISM: lines and
  bands --- ISM: molecules --- supernovae: general --- supernovae:
  individual (SN\,2012ap)}

\section{Introduction}

One of the long unsolved problems in optical and infrared astronomy is
the nature of the diffuse interstellar bands (DIBs). The DIBs
represent more than 400 absorption features observed in optical and
near-infrared spectra that are typically narrow (full width at
half-maximum intensity [FWHM] $< 1$\,\AA) and weak (less than 5\%
below the continuum), with central wavelengths that do not correspond
with any known atomic or molecular species
\citep{Herbig95,Hobbs09,Geballe11}. They were first noticed in stellar
spectra by \citet{Heger22}. \citet{Merrill34} subsequently uncovered a
number of DIBs as ubiquitous interstellar features and their nature
has been an enduring subject of speculation.

It is now well established that sources (or ``carriers'') of the DIBs
are found in the interstellar medium (ISM). DIB features remain
stationary in spectroscopic binaries, and there are rough correlations
between extinction and \ion{Na}{1}\,D column density with the
intensity of DIB features \citep{Herbig95}. Searches for DIBs in
circumstellar shells have generally reported null detections or
results that cannot distinguish whether the absorption arises in
circumstellar material or the intervening ISM \citep{Snow72,Luna08}.

\begin{figure*}[htp!]
\centering

\includegraphics[width=0.65\linewidth]{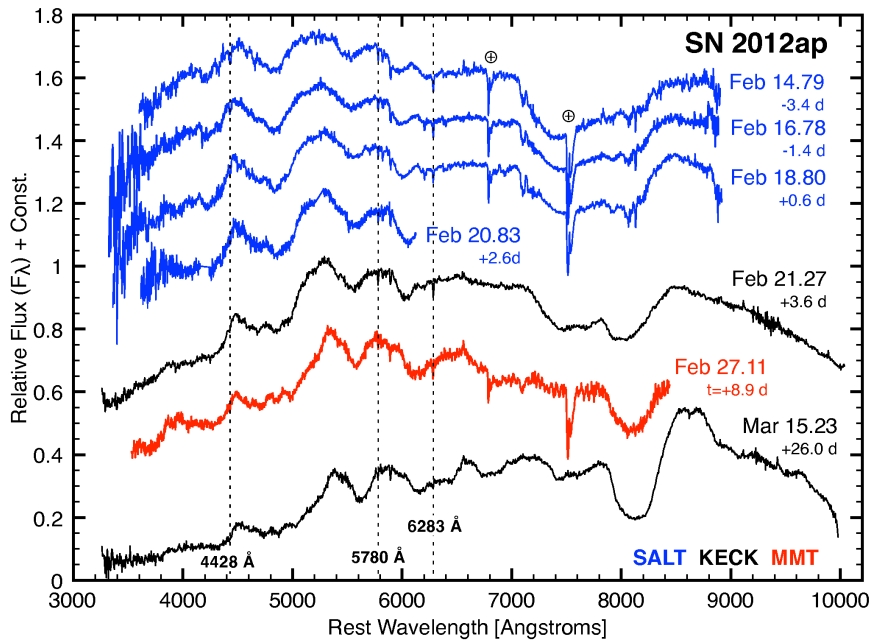}

\caption{Optical spectra of the broad-lined Type Ic SN\,2012ap during
  its first $\sim 40$\,d after explosion. Three prominent DIBs around
  4428\,\AA, 5780\,\AA, and 6283\,\AA\ are highlighted with dashed
  lines. SALT and Keck data have a 4--6\,\AA\ FWHM resolution. MMT
  data have 7\,\AA\ FWHM. Symbols show telluric absorptions in the
  SALT spectra that have not been corrected. Spectra have been
  corrected for a redshift of z = 0.0121 measured from narrow
  \ion{H}{2} region lines of [\ion{O}{3}] \dlambda 4959, 5007,
  H$\alpha$, and [\ion{N}{2}] \dlambda 6548, 6583 observed near the
  location of the supernova.}

\label{fig:spectra-wide}
\end{figure*}

\citet{Merrill34} was the first to suspect dust grains and/or
molecules as possible carriers of the DIBs.  After nearly a century of
observational, theoretical, and experimental work, these two original
suggestions have remained the primary candidates, occasionally
swapping in popularity (see \citealt{Sarre06}, and references
therein). Current research favors multiple carriers produced by a mix
of fairly large and complex carbon-based (``organic'') polyatomic
molecules composed of cosmically abundant elements such as H, C, O,
and N. There has been considerable investigation of polycyclic
aromatic hydrocarbons (PAHs) as DIB carriers, but as yet no firm
associations between PAH species and DIB features have been found (see
\citealt{Cox11} for a recent review).

Insights into the chemical and physical properties of DIB carriers has
come from study of their behavior in different interstellar
environments, especially extragalactic ones.  Most studies have
focused on nearby star systems including the Magellanic Clouds and M31
\citep{Cox07,Cordiner08a}. Outside of limited work with quasars (e.g.,
\citealt{Ellison08}), only supernovae (SN) have been bright enough to
probe DIBs beyond the Local Group (see, e.g., \citealt{Cox08}). In
general, extragalactic studies have shown that DIB carrier abundances
can be similar to Galactic values, though systematic differences
sometimes exist.

In this {\it Letter} we report on recent observations of a broad-lined
Type Ic SN that exhibits some of the strongest DIBs ever detected in
an extragalactic source. These absorptions undergo changes in
intensity over relatively short timescales in a manner that suggests
that the SN explosion interacted with local carriers of DIBs. We
conclude that moderate-resolution spectra of SN obtained shortly after
outburst may provide a new and powerful probe of DIBs and offer clues
about the progenitor systems of these explosions.

\section{Results}

\subsection{Discovery and Classification}

SN\,2012ap was first detected by the Lick Observatory Supernova Search
at coordinates $\alpha(2000.0)$ = 05$^h$00$^m$13$\fs$72 and
$\delta(2000.0)$ = $-03\degr$20$\arcmin$51$\farcs$2 in the face-on
galaxy NGC 1729 ($d \approx 43.1$ Mpc; \citealt{Springob09}) on Feb. 10.23
UT \citep{Jewett12}. The SN is located in the
outskirts of the host galaxy some 7.1\,kpc in projection from the
nucleus in a region with no obvious star formation.

The first reports of optical spectra of SN\,2012ap classified it as a
Type Ib/c SN similar to SN 2008D not long after explosion
\citep{Xu12}. This prompted extensive follow-up observations by our
group that included optical spectra obtained with the 10\,m Southern
African Large Telescope (SALT) using the Robert Stobie Spectrograph
(RSS; \citealt{Burgh03}), the 10\,m Keck-I telescope using the Low
Resolution Imaging Spectrometer \citep[LRIS;][]{Oke95}, and the MMT
6.5\,m telescope using the Blue Channel spectrograph
\citep{Schmidt89}. The spectra shown in Figure~\ref{fig:spectra-wide}
are part of a larger dataset (Milisavljevic et al., in prep.).

Unlike SN\,2008D, which transitioned to a Type Ib SN exhibiting
conspicuous \ion{He}{1}, spectra of SN\,2012ap obtained weeks later
continued to show broad features associated with ejecta traveling
$\sim 2 \times 10^4$\,\kms. \citet{Milisavljevic12} reported that
these later spectra were similar to those observed in broad-lined
SN~Ic such as SN\,1998bw and SN\,2002ap $\sim 1$--2 weeks after
maximum light (see Fig.~\ref{fig:spectra-wide}). Further examination
shows that the later spectra of SN\,2012ap also resemble those of
SN\,2009bb, a SN~Ib/c that had a substantial relativistic outflow
powered by a central engine \citep{Soderberg10,Pignata11}.

\begin{figure*}[htp!]
\centering
\includegraphics[width=\linewidth]{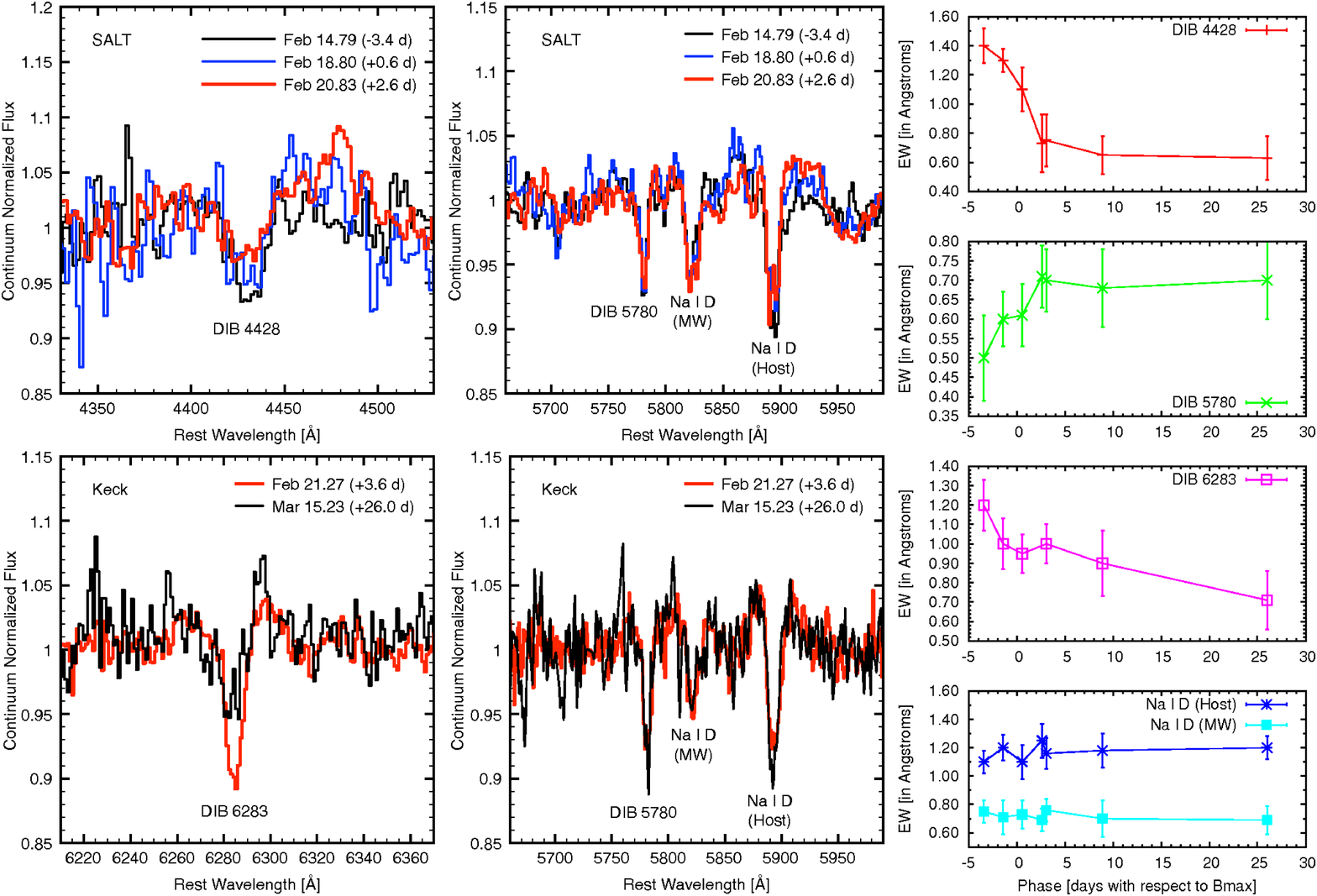}

\caption{Enlarged DIB absorption features of SN\,2012ap. The left and
  middle columns show select epochs to illustrate possible changes in
  EW, and the right column shows all EW measurements of narrow
  absorptions. Phase is with respect to $B$-band maximum on Feb 18.2
  UT.  We removed the global shape of the underlying continuum by
  smoothing the spectra with a running boxcar of width 50\,\AA\ and
  then subtracting the smoothed version. The EW of the \ion{Na}{1}\,D
  lines of the host galaxy shows no measurable change. To further
  illustrate that the changes are not related to instrumental setups,
  spectra from the same the spectrograph using identical
  configurations are plotted.}

\label{fig:spectra-narrow}
\end{figure*}

\subsection{Strong DIB Features}

Superimposed on the broad-lined Type Ic features of SN\,2012ap are
conspicuous absorptions with equivalent widths (EWs) $\ga 1$\,\AA\
associated with DIBs at the rest wavelength of the host galaxy. The
DIB features are strongest at 4428\,\AA, 5780\,\AA, and 6283\,\AA,
which are the wavelengths of well-known DIBs typically seen in stellar
spectra \citep{Herbig95}. In Figure~\ref{fig:spectra-narrow} we
display enlarged regions around these features. Not shown is another
possible DIB detection near 6203\,\AA\ that may be contaminated by an
OH telluric line at an observed wavelength of 6280\,\AA.

The central wavelengths of these DIBs do not change with time, but the
intensities do exhibit measurable changes that are not uniform across
different features (Fig.~\ref{fig:spectra-narrow}, right column). The
EW of DIB $\lambda$4428 decreased by $0.77\pm0.25$\,\AA\ over $\sim
10$ days and DIB $\lambda$6283 decreased by $0.49\pm0.28$\,\AA\ over
$\sim 30$ days. The DIB $\lambda$5780 feature, on the other hand,
shows a weak but measurable increase of $\la 0.2$\,\AA\ over $\sim 10$
days. The \ion{Na}{1}\,D line at rest with respect to the host shows
negligible change. The \ion{Na}{1}\,D line associated with foreground
Milky Way extinction shows no change, as expected.

\section{Discussion}

\subsection{SN Interaction with DIB Carriers}

The DIB absorptions seen in the spectra of SN\,2012ap are
among the strongest extragalactic detections ever reported. Detections
of extragalactic DIBs at this distance are rare and thus
interesting as they allow one to compare Galactic ISM chemical
properties with extragalactic ones. However, what is unique and most
informative about the spectra of SN\,2012ap is that the DIB absorption
strengths {\it change with time} and that {\it the changes are not
  uniform across different DIB features} (see
Fig.~\ref{fig:spectra-narrow}).

Various types of interaction between the SN and DIB carrier material
may explain the observed changes (see, e.g., \citealt{Patat10}). We
favor the scenario that the carrier material is nearby and the SN is
actively interacting with it. This interaction can take many
forms. Photons may modify or destroy carrier material via ionization
and/or dissociation. If extremely nearby, the forward blast wave
initiated by the explosion and traveling with velocity $\sim 0.4c$
(Chakraborti et al., in prep.) will disrupt molecules and dust grains
within a $\sim 0.01$\,pc radius in the first 30 days.

\begin{figure*}[htp!]
\centering
\includegraphics[width=0.75\linewidth]{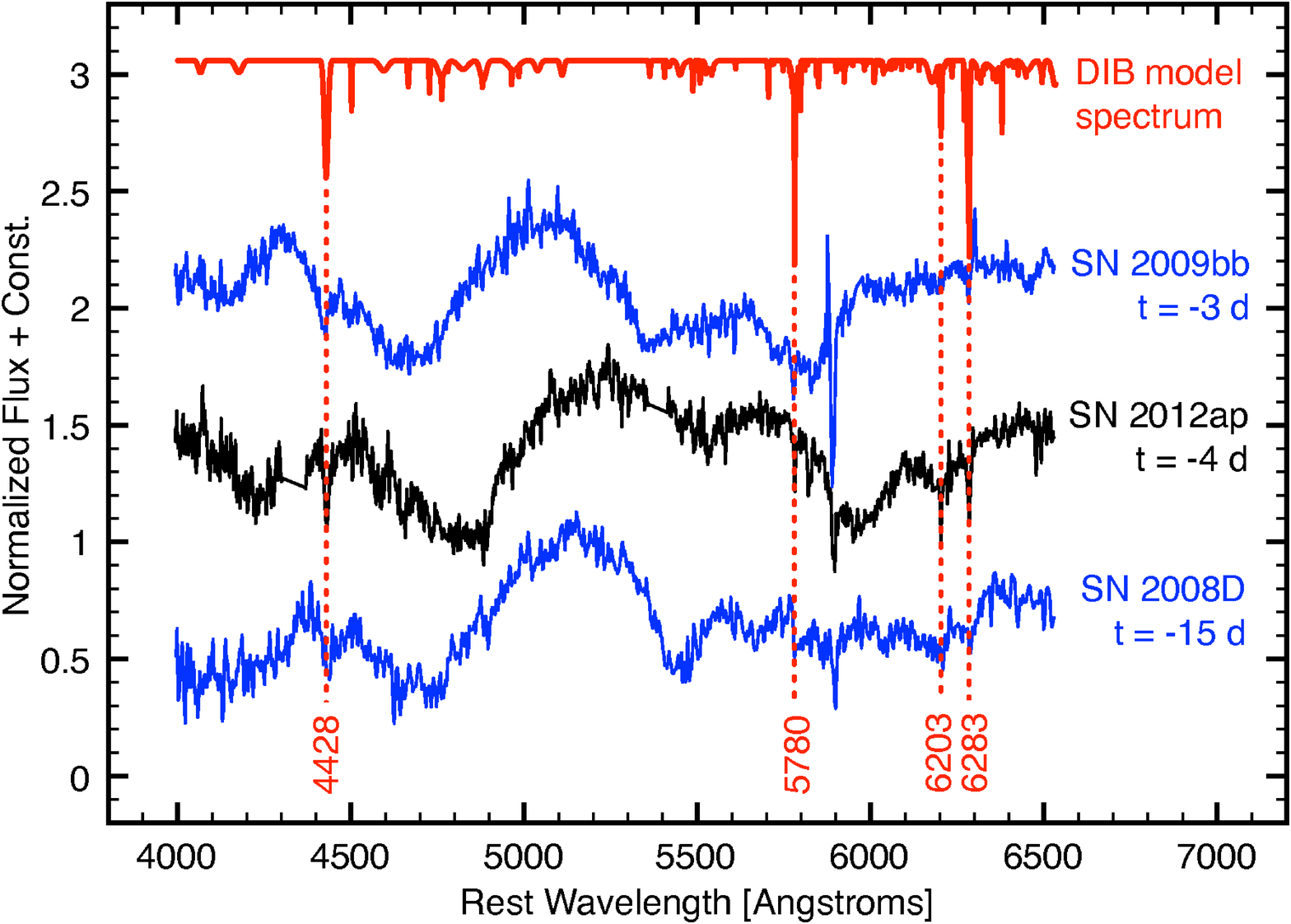}

\caption{Model spectrum of known DIB absorption features compared to
  early-time spectra of SN 2012ap, SN 2009bb \citep{Pignata11}, and SN
  2008D \citep{Modjaz09}. SN 2009bb and SN 2008D have been corrected
  for redshifts of z = 0.010 and z = 0.007, respectively. All SN
  exhibit conspicuous absorptions having central wavelengths of
  well-known DIBs highlighted with vertical dashed lines. Time is with
  respect to maximum light. The model DIB spectrum was created from a
  catalog retrieved online at
  http://leonid.arc.nasa.gov/DIBcatalog.html. }

\label{fig:examples}
\end{figure*}

\subsection{Physical Constraints on the DIB Carriers}

SN\,2012ap peaked in the $B$ band on Feb 18.2 UT (Milisavljevic et
al., in prep.), implying that the SN flux increased and then decreased
at optical wavelengths as the intensities of DIB $\lambda$4428 and
$\lambda$6283 became weaker and DIB $\lambda$5780 became
slightly stronger. This behavior is consistent with active interaction
wherein separate DIB carriers differing in robustness and/or location
are affected by the SN independently.

Using the time evolution of the blackbody temperature and total
luminosity derived from photometry data (Milisavljevic et al., in
prep.), we estimated the UV flux in the 5--50\,eV spectral range
as a function of distance and time from SN\,2012ap. To estimate the
lifetimes of molecular material in this radiation field, we
approximated the photoabsorption cross sections in this frequency
range for small neutral molecules \citep{Gallagher88} and PAHs
\citep{Verstraete90,Jochims96}, calculated the photoabsorption rate,
and assumed that all absorption events lead to ionization or
dissociation. Because these frequencies are above the peak of the
blackbody curve, the absorption rates are highly sensitive to the
ionization potential (IP) of a molecule, and the shape and size of its
cross section.

The inferred lifetimes vary by several orders of magnitude, but within
a distance of $\sim 0.01$\,pc, at peak luminosity all but the smallest
neutral molecules are expected to have lifetimes much less than one
day. Within this distance, the population of most neutral species will
be rapidly depleted unless their formation from the breakdown of
larger material is even more rapid. Cations, owing to their higher
IPs, are estimated to have lifetimes on the order of days under the
same conditions.

In this context, it is interesting to note that the timescale for the
increase in DIB $\lambda$5780 is comparable to that of the decay in
DIB $\lambda$4428, possibly suggesting that the DIB $\lambda$5780
carrier is a photoproduct of the DIB $\lambda$4428 carrier. In
constrast, the decay in DIB $\lambda$6283 occurs over a longer
timescale, suggesting the carrier is either more photostable or is
more extended. The ratio of the strength of DIB $\lambda$5780 to DIB
$\lambda$5797 (the latter is not detected toward SN\,2012ap) is
positively correlated with increasing UV radiation environments
\citep{Vos11}. The increase in strength of DIB $\lambda$5780 in these
observations suggests that this trend continues to very extreme UV
environments.

Fullerenes have been proposed as DIB carriers, and are significantly
more stable against dissociation by UV radiation than smaller
molecules, typically requiring energies of more than 10\,eV for
dissociation \citep{Diaz-Tendero03}.  This increased dissociation
energy might allow fullerenes to survive longer in the radiation
environment around SN\,2012ap. Using the photoabsorption cross section
of C$_{60}$ as a representative case \citep{Berkowitz99}, we estimate
that neutral fullerenes (IP $\approx 7$\,eV) near SN\,2012ap will be
rapidly ionized, but fullerene cations (IP $\sim 11$\,eV) should have
lifetimes of order days. The fact that the observed changes in EW of
these DIB features occur on the timescale of days in such an intense
UV field suggests the carriers are fairly robust to ionization and
dissociation (particularly DIB $\lambda$5780), consistent with small
cations or charged fullerenes.

\subsection{Implications of a DIB--SN Subtype Correlation}

Two other core-collapse SN in the literature exhibit conspicuous DIBs
in low-resolution spectra, and we examined their archival data: the
Type Ib SN\,2008D with spectra published by \citet{Modjaz09}, and the
broad-lined Type Ib/c SN\,2009bb published by
\citet{Pignata11}. Figure~\ref{fig:examples} shows early-time spectra
of these objects, with conspicuous DIB features highlighted. Although
the relatively low spectral resolutions and limited temporal sampling
prevent detailed analyses of these additional objects, the archival
spectra suggest that some DIB features seen in these other SN have
both narrow and broad components and that they may vary as they do
SN\,2012ap.

All three SN exhibited broad spectral features associated with ejecta
moving at high velocities ($\ga 2 \times 10^4$\,\kms) within weeks of
explosion and all were observed to have a color excess $E(B-V) \ga
0.5$ mag that implies substantial extinction
(\citealt{Soderberg08,Modjaz09,Pignata11}; Milisavljevic et al., in
prep.). SN\,2012ap and SN\,2009bb share similar explosion parameters
of estimated ejecta mass ($\sim 2$--4\,M$_{\odot}$), $^{56}$Ni mass
($\sim 0.2$\,M$_{\odot}$), and explosion kinetic energy ($\sim 1.5
\times 10^{52}$\,erg). On the other hand, SN\,2008D is different in
that its broad lines disappeared within weeks as it transitioned to a
SN~Ib and its explosion energy ($\sim 1.5$--6 $\times 10^{51}$\,erg;
\citealt{Soderberg08,Tanaka09}) is lower than those of SN\,2012ap and
SN\,2009bb.

Chance alignments between DIB carrier-rich molecular clouds and these
SN are possible. However, given that the three SN with conspicuous DIB
absorptions examined in the literature are spectroscopically similar,
it may be that the SN progenitor systems are related to the sources of
the DIBs. If true, the carrier material responsible for the observed
DIB absorptions in these SN should lie fairly close to the explosion
site and could be associated with mass loss from the progenitor star.

Mass loss in massive stars is influenced by a number of factors
including the strength of their winds, rotation, the presence of a
binary companion, possible eruptive mass-loss episodes, and
environmental metallicity \citep{Chiosi86,Humphreys94,Nugis00}.  To
investigate what role metallicity might play in linking the three SN,
the relative strengths of narrow lines from coincident host-galaxy
emission at the site of SN\,2012ap were measured using the method
described by \citet{Sanders12}. From the N2 diagnostic of
\citet{PP04}, we measure an oxygen abundance $\log({\rm O/H})+12=8.79$
with uncertainty 0.06 dex. Adopting a solar metallicity of
$\log({\rm O/H})_{\odot} + 12 = 8.69$ \citep{Asplund05}, our measurement
indicates that SN\,2012ap exploded in an environment of around solar
metallicity that lies in between the metallicity estimates of
SN\,2009bb (1.7--3.5\,Z$_{\odot}$; \citealt{Levesque10}) and SN\,2008D
(0.5--1\,Z$_{\odot}$; \citealt{Soderberg08}). Considering broad-lined
SN Ic are typically found in environments of subsolar metallicity
\citep{Kelly12,Sanders12}, the metallicity of these three SN
is somewhat anomalous. However, these objects were discovered by
surveys targeting high-mass metal-rich galaxies, so this weak trend
may be influenced by an observational bias.

A handful of reports connect strong DIB features observed in a narrow
subset of mass-losing stars with circumstellar shells (e.g.,
\citealt{Tug81,Cohen87}). The circumstellar material is often
nitrogen-rich and the strength of the associated DIB features may vary
\citep{Heydari93}.  \citet{LeBertre93} identified Wolf-Rayet (WR)
stars of the WN subtype and luminous blue variable (LBV) stars
enriched in nitrogen as candidate objects with circumstellar shells
containing DIB carriers, and proposed that nitrogen could act either
as a constituent of the DIB carriers or as a catalyst for their
production.

It is intriguing that families of WR and LBV stars may be associated
with DIB features. WR stars are suspected progenitors of SN~Ib/c
\citep{Gaskell86}, and have been implicated for SN\,2008D and
SN\,2009bb \citep{Soderberg08,Modjaz09,Soderberg10,Pignata11}.
Although LBVs are not widely believed to be the direct progenitors of
SN~Ib/c, WR stars can evolve from a prior LBV phase
\citep{Conti76}. These stars exhibit varying degrees of asymmetric
mass loss (see, e.g., \citealt{Nota95}), thus an observer's line of
sight with respect to a circumstellar disk could be an important
factor in explaining why strong DIB detections like those reported
here are rare.

Finally, we note that varying strength in narrow absorption lines
attributable to interaction between a SN and a local environment has
recently been recognized in a growing number of cases, with
significant implications for the nature of the progenitor systems
(e.g., \citealt{Patat07,Blondin09,Dilday12}). However, those reports
have been for \ion{Na}{1}\,D, \ion{Ca}{2}, H$\alpha$, \ion{He}{1}, and
\ion{Fe}{2} lines with line-of-sight blueshifted velocities of $\la
100$\,\kms\ originating from circumstellar material around Type Ia
SN. This is not the same as what is being observed in the
core-collapse SN\,2012ap, where the DIB features are near zero
velocity and are associated with a carrier material having radically
different physical properties.

\section{Conclusions}

The broad-lined Type Ic SN\,2012ap exhibits DIB absorptions that are
among the strongest ever detected in an extragalactic object.  The DIB
features centered around 4428\,\AA, 5780\,\AA, and 6283\,\AA\ undergo
changes in EW over relatively short timescales ($t<30$ days)
indicative of interaction between the SN and DIB carriers.  Similar
absorptions observed in archival spectra of two additional SN suggests
that SN\,2012ap may belong to a subset of energetic SN~Ib/c that
exhibit changes in conspicuous DIB absorption features. If true, this
correlation is consistent with the DIB carrier-rich material being
located close to the explosion, fairly resistent to the strong UV
field, and potentially associated with mass loss of the progenitor
star.

Our data with 4--7~\AA\ resolution that monitored the spectral
evolution of SN\,2012ap during its rise and fall in flux was on the
cusp of detection for this uniquely strong source of DIB absorptions.
Only the broadest DIB features known to have FWHM widths of
approximately $2-12$ \AA\ were observed in our data set. Thus,
multi-epoch observations of SN with spectral resolutions of $\leq
1$~\AA\ beginning within days of explosion could uncover the presence
of a larger family of DIB features. Such observations would be much
more sensitive to possible variations in \ion{Na}{1} absorption
strength, as well as detect possible subtle changes in the velocities
of the NaI/DIB features. Observed in this way, SN with DIB
absorptions have the potential to reveal unique information about
mass-loss environment of their progenitor systems and probe DIB
carriers in new ways that can bring us closer to understanding the
their nature.

\acknowledgements

We thank an anonymous referee for a helpful, detailed and critical
reading of the paper. T. Snow kindly provided comments on an early
draft of the paper. P.\ Massey provided insightful
comments. G. Pignata, S. Valenti, D. Malesani, and G. Leloudas shared
archival spectra that were examined. Many of the observations reported
in this paper were obtained with the Southern African Large
Telescope. Additional data presented herein were obtained at the W.~M.
Keck Observatory, which is operated as a scientific partnership among
the California Institute of Technology, the University of California,
and NASA; the observatory was made possible by the generous financial
support of the W.~M.\ Keck Foundation. A.  Miller, P. Nugent, and
A. Morgan helped obtain the Keck observations. Some observations also
came from the MMT Observatory, a joint facility of the Smithsonian
Institution and the University of Arizona. Support was provided by the
David and Lucile Packard Foundation Fellowship for Science and
Engineering awarded to A.M.S.  J.M.S. is supported by an NSF Astronomy
and Astrophysics Postdoctoral Fellowship under award
AST-1302771. T.E.P. thanks the National Research Foundation of South
Africa. R.P.K. and J.C.W. are grateful for NSF grants AST-1211196 and
AST-1109801, respectively.  A.V.F. and S.B.C. acknowledge generous
support from Gary and Cynthia Bengier, the Richard and Rhoda Goldman
Fund, the Christopher R. Redlich Fund, the TABASGO Foundation, and NSF
grant AST-1211916.  K.N.C. has been supported by a CfA Postdoctoral
Fellowship from the Smithsonian Astophysical Observatory. This paper
made extensive use of the SUSPECT database
(\texttt{http://www.nhn.ou.edu/$\sim$suspect/}).

\end{document}